\documentclass[twocolumn,showpacs,preprintnumbers,amsmath,amssymb]{revtex4}
\usepackage{dcolumn}
\usepackage{bm}
\usepackage{graphicx,subfigure,xcolor}

\usepackage{amsmath}
\usepackage{amssymb}
\usepackage{amsfonts}
\usepackage{latexsym}

\newcommand{\wo}{\omega_0}
\newcommand{\bd}{b^\dagger}
\newcommand{\ad}{a^\dagger}
\newcommand{\Hil}{\mathcal H}
\newcommand{\D}{{\mathcal D}}
\newcommand{\F}{{\cal F}}
\newcommand{\G}{{\cal G}}
\newcommand{\1}{1 \!\! 1}

\begin{document}

\title{Non-Hermitian Hamiltonian for a modulated Jaynes-Cummings model with ${\mathcal PT}$ symmetry}

\author{Fabio Bagarello\mbox{$^{1}$}, Margherita Lattuca\mbox{$^{2}$}, Roberto Passante\mbox{$^{2}$}, Lucia Rizzuto\mbox{$^{2}$}, Salvatore Spagnolo\mbox{$^{2}$}}
\affiliation{\mbox{$^{1}$}Dipartimento di Energia, Ingegneria dell'Informazione e Modelli Matematici, Universit\`a degli Studi di Palermo, I-90128 Palermo, Italy, and INFN, Sezione di Torino, Italy\\
\mbox{$^{2}$}Dipartimento di Fisica e Chimica, Universit\`{a} degli Studi di Palermo and CNISM, Via Archirafi 36, I-90123 Palermo, Italy}

\email{roberto.passante@unipa.it}


\begin{abstract}
We consider a two-level system such as a two-level atom, interacting with a cavity field mode in the rotating wave approximation, when the atomic transition frequency or the field mode frequency is periodically driven in time. We show that in both cases, for an appropriate choice of the modulation parameters, the state amplitudes in a generic $n${-}excitation subspace obey the same equations of motion that can be obtained from a \emph{static} non-Hermitian Jaynes-Cummings Hamiltonian with ${\mathcal PT}$ symmetry, that is with an imaginary coupling constant. This gives further support to recent results showing the possible physical interest of  ${\mathcal PT}$ symmetric non-Hermitian Hamiltonians. We also generalize the well-known diagonalization of the Jaynes-Cummings Hamiltonian to the non-Hermitian case in terms of pseudo-bosons and pseudo-fermions, and discuss relevant mathematical and physical aspects.
\end{abstract}

\maketitle

\section{Introduction}
\label{sec:Introduction}

Non-Hermitian Hamiltonians have been frequently used in physics to describe dissipative effects in concrete systems in many different fields of physics, in particular in quantum optics
\cite{BR03,Peng14}, or for describing decaying states \cite{SW72}.
Recently, there has been a renewed interest in non-Hermitian Hamiltonians after it has been shown that non-Hermitian Hamiltonians with ${\mathcal PT}$ (parity-time) symmetry can have a real eigenvalues \cite{BB98,BBJ02,Bender07}. The same happens for non-Hermitian  but pseudo-symmetric Hamiltonians, \cite{mosta2002}, where the ${\mathcal PT}$ symmetry is replaced by a more abstract condition.
In a recent paper, it has been shown that physical systems such as optical lattices with some external modulation of their parameters can be described by a Lee-Friedrichs Hamiltonian in the ghost regime, that is when the coupling constant is imaginary and therefore the Hamiltonian is not {Hermitian}\cite{BBCW05,LV12}.
In this paper we consider another physical system, of interest in quantum optics, with a time dependence of a physical parameter that can be simulated by a time-independent non-Hermitian Hamiltonian with an imaginary coupling constant.

We consider a two-level system, a two-level atom for example, interacting with a boson field, for example a single cavity mode of the electromagnetic field, when one parameter of the system is periodically modulated in time. Specifically, we investigate two cases: a modulation of the transition frequency of the two-level system and a modulation of the frequency of the cavity mode. Cavities with a conducting wall subjected to a periodic motion and optically modulated cavities have been extensively considered in the literature, in particular in the framework of the dynamical Casimir effect \cite{FD76,FC11}. We consider the interaction of the two-level atom with a single cavity mode in the rotating wave approximation, and thus the system is described by a Jaynes-{Cummings}(JC) Hamiltonian \cite{JC63,CTDRG92,CPP95}, with one time-dependent parameter (atomic transition frequency or cavity field frequency). Our system is thus described by a Hermitian time-dependent JC Hamiltonian, and we obtain the differential equations for the state amplitudes in each subspace characterized by a well-defined number of total (atom plus field) excitations. This is possible because our Hamiltonian commutes with the total excitation-number operator. We show that, under an appropriate choice of the parameters characterizing the system's modulation and after a time-average, these equations are identical to those that can be obtained by a \emph{static} JC Hamiltonian with an imaginary coupling constant, that is a non-Hermitian JC Hamiltonian. This Hamiltonian, however, still has a ${\mathcal PT}$ symmetry. This result gives a physical meaning and importance in considering a non-Hermitian Hamiltonian for a matter-field interacting system, specifically a fermion field linearly coupled to a boson field. This is an extension of an analogous recent result considering a modulated Lee-Friedrichs Hamiltonian (that is in the one-excitation subspace), where light transport in a semi-infinite waveguide lattice with a nonlinear modulation was considered and equivalence to a non-Hermitian Lee-Friedrichs Hamiltonian in the ghost regime was shown \cite{LV12}.
Due to the physical relevance of a non-Hermitian JC Hamiltonian, we then generalize the well-known diagonalization  of the JC Hamiltonian by a unitary transformation to the non-Hermitian case in terms of pseudo-boson \cite{Bagarello11,BL13} and pseudo-fermion operators \cite{Bagarello13},  which have been recently introduced for fermionic and bosonic systems separately.
{These pseudo-Hermitian operators arise from a special deformation of canonical bosonic and fermonic commutation relations \cite{Bagarello11, Bagarello13}.
The pseudo-Hermitian Hamiltonian we obtain is physically equivalent to the non Hermitian ${\mathcal PT}$ symmetric Hamiltonian and can be diagonalized exactly, generalizing the procedure used in the Hermitian case.
Thus, our} results extend the definition of pseudo-bosons and pseudo-fermions to the case of a fermion-boson interacting system such as the Jaynes-Cummings model, allowing us a mathematically rigorous treatment of our non-Hermitian interaction Hamiltonian.
{The relevance and application of the formalism of the pseudo-Hermitian operators in the framework of open quantum system or in quantum optics has been recently discussed in the literature (see for example \cite{Mostafazadeh10} and reference therein)}

This paper is organized as follows. In Sec. \ref{sec:JCModel} we introduce our ${\mathcal PT}$ symmetric Jaynes-Cummings model with a modulation of the atomic or field frequency. In Sec. \ref{sec:JCSolution} we consider the equations of motion for the state of the interacting system in each subspace characterized by a well-defined number of excitations, and show that, for appropriate choices of the system parameters, they are equivalent to a static non-Hermitian JC Hamiltonian. In Sec. \ref{sec:nonHJC}, motivated by our previous analysis, we diagonalize the non-Hermitian JC Hamiltonian through an appropriate transformation operator, generalizing the well-known diagonalization for the Hermitian case, in terms of pseudo-boson and pseudo-fermion operators, and discuss some mathematical and physical aspects. Sec. \ref{sec:Concl} is finally devoted to our concluding remarks.

\section{The Model}
\label{sec:JCModel}

A two-level atom interacting with a single mode of the electromagnetic field in the rotating wave approximation (RWA), can be described by the Jaynes-Cummings model \cite{JC63,CTDRG92,CPP95}. This Hamiltonian is frequently used in quantum optics, in particular for studying non-perturbative effects in strong coupling regimes \cite{Kimble98,RBH01}. The relative Hamiltonian can be written as
\begin{equation}
\label{JCHamiltonian}
H = H_0+V= \hbar \wo \left( \bd b -\frac 12 \right) + \hbar \omega \ad a + g_s \left( a\bd +  \ad b \right) \, ,
\end{equation}
where the annihilation and creation operators $\bd \, , b$ refer to the two-level system with transition frequency $\wo$ and satisfy standard fermionic anti-commutation relations, $\ad \, , a$ refer to the field mode with frequency $\omega$ and satisfy standard bosonic commutation relations, and $g_s$ is a coupling constant that in this moment we assume to be real, so that the Hamiltonian \eqref{JCHamiltonian} is clearly Hermitian. The JC Hamiltonian is the simplest Hamiltonian model describing the coupling between a fermionic and a bosonic field, and it can be diagonalized exactly. In fact, it is  straightforward to verify that it commutes with the operator giving the total number of excitations (atom plus field)
\begin{equation}
\label{NumberExc}
{\cal N}= \bd b + \ad a  \, .
\end{equation}
Thus the Hamiltonian can be diagonalized independently in each subspace characterized by a well-defined number of excitations. In the $n$-excitation subspace, a basis is given by the two states $\mid n,0 \rangle$ and $\mid n-1,1 \rangle$, where the first element in the states indicates the number of excitations in the bosonic sector, and the second one that in the fermionic sector. The first state has $n$ quanta in the field mode and the two-level system is in its ground state, while in the second state there are $n-1$ quanta in the field and the two-level system is in its excited state.
In the $n$-excitation subspace, the eigenstates and the eigenvalues of the JC Hamiltonian are well known \cite{JC63,CTDRG92,CPP95}
\begin{eqnarray}
\label{JCeigen}
\mid u_n^{(+)} \rangle &=& \cos \frac \theta 2 \mid n,0 \rangle - \sin \frac \theta 2 \mid n-1,1 \rangle \, , \nonumber \\
\mid u_n^{(-)} \rangle &=& \cos \frac \theta 2 \mid n-1,1 \rangle + \sin \frac \theta 2 \mid n,0 \rangle \, , \nonumber \\
E_n^{(\pm )} &=& \left( n-\frac 12 \right) \hbar \omega \pm \Delta /2 \, ,
\end{eqnarray}
where
\begin{eqnarray}
\label{Defs}
\sin \frac \theta 2 &=& \frac 1{\sqrt{2}} \left( 1 + \frac \delta \Delta \right)^{1/2} \, , \nonumber \\
\cos \frac \theta 2 &=& -\frac 1{\sqrt{2}} \left( 1- \frac \delta \Delta \right)^{1/2} \, , \nonumber \\
\Delta &=& \left( \delta^2 +4\mid g_s \mid^2 n \right)^{1/2} \, ,
\end{eqnarray}
with $\delta = \hbar ( \omega_0 - \omega )$ the detuning in energy between the two-level system and the field mode. The dressed states $\mid u_n^{(\pm )}\rangle$ are entangled atom-field states, and the interaction removes the degeneracy present at resonance ($\omega = \omega_0$) in each subspace with $n \neq 0$ (the ground state is not degenerate and it is not shifted in energy by the interaction). Other analogous models frequently used to deal with a discrete system interacting with field modes, and that can be exactly diagonalized by {Bogoliubov}-type transformations, include an harmonic oscillator interacting with a set of independent harmonic oscillators (see for example \cite{PRSTP12}).

The relation between bare and dressed states of the JC model can be also expressed in terms of the unitary operator \cite{CPP95,CP80}
\begin{equation}
\label{Toperator}
T= \exp\left\{-\hat{\theta} \left( 4 {\cal N}\right)^{-1/2}\left(  a\bd - \ad b \right)\right\} \, ,
\end{equation}
where $\hat{\theta}$ is an operator defined as
\begin{equation}
\label{ThetaOperator}
\sin \hat{\theta} = -\left( 4\mid g_s \mid^2 {\cal N}\right)^{1/2} \hat{\Delta}^{-1}\, , \, \, \,
\cos \hat{\theta} = - \delta \hat{\Delta}^{-1} \, ,
\end{equation}
with
\begin{equation}
\label{DeltaOperator}
\hat{\Delta} = \left( \delta^2+ 4\mid g_s \mid^2 {\cal N}\right)^{1/2} \, .
\end{equation}
Of course, if $\delta\neq0$, the operator $\delta^2+4|g_s|^2 {\cal N}$ is self-adjoint and strictly positive. Then it admits a positive square root, $\hat{\Delta}$, which is surely invertible. On the other hand, strictly speaking, ${\cal N}^{-1/2}$ is not well defined, since zero is one of the eigenvalues of $\cal N$. However, the inverse of $\cal N$ (and of its square root) does exist in any subspace of $\Hil$ which does not contain the vacuum of $a$ and $b$, which are the only subspaces relevant in our analysis.

The Hamiltonian $H$, if expressed in terms of the transformed operators, assumes a diagonal form,
\begin{equation}
\label{DiagonalH}
H=\hbar \omega \bar{a}^\dagger \bar{a} + \left( \hbar \omega - \hat{\Delta} \right) \left( \bar{b}^\dagger \bar{b} -\frac 12 \right) \, ,
\end{equation}
where $\bar{O}=TOT^{-1}$ is the transformed operator of $O$ (${\cal N}$ is invariant under the transformation $T$: $\bar{\cal N}=T{\cal N}T^{-1}={\cal N}$). Then the dressed eigenstates \eqref{JCeigen} can be obtained from the corresponding bare ones by application of the unitary operator $T$.
Later on in this paper we shall extend this transformation to the case of a non-Hermitian Jaynes-Cummings Hamiltonian model, specifically when the coupling constant
$g_s$ in \eqref{JCHamiltonian} is assumed complex. This is a situation occurring in the renormalized Lee-Friedrichs model in the so-called ghost regime, that is when the bare coupling constant is larger than a critical value \cite{LV12,Lee54,Schweber61,BBCW05}. The Lee-Friedrichs model \cite{Lee54} has some similarity with the Jaynes-Cummings model we are considering, main difference being that the Lee-Friedrichs model contains an infinite (continuous) set of field modes, and it is usually restricted to the one-excitation subspace only. In the case of the Lee model with imaginary coupling constant (ghost regime), it has been shown that it still has a ${\mathcal PT}$ (parity-time) symmetry  \cite{LV12,BBCW05}, and that it can be usefully used to describe the light transport in an infinite waveguide lattice with a time-dependent nonlinearity \cite{LV12,Longhi10}. Also, it has been shown that the Hamiltonian of quantum electrodynamics becomes non-Hermitian when the unrenormalized electric charge is taken to be imaginary, but the Hamiltonian is ${\mathcal PT}$ symmetric, and the time evolution unitary, if the quadripotential transforms as a {pseudovector} \cite{BCPMS05}.

Following similar lines of the discussion in \cite{BBCW05} for the Lee model, we can show that also the JC Hamiltonian \eqref{JCHamiltonian} is ${\mathcal PT}$ symmetric even if the coupling constant $g_s=ig$ becomes imaginary (and thus the Hamiltonian is not Hermitian), provided the real quantity $g$ is such that
${\mathcal PT}g{\mathcal PT}=-g$ .
In the next section, we will show that the non-Hermitian JC Hamiltonian proves very useful to describe the properties of a two-level system interacting with a field cavity mode, when a periodic modulation of the frequency of the two-level system or of the cavity mode is given to the system. In the first case such a modulation can be experimentally obtained by Stark shift through the interaction of a two-level atom with an external periodic electric field; in the second case it can be obtained through a mechanical oscillation of a cavity wall, which could be also obtained by a dynamical mirror, that is a wall whose dielectric properties are rapidly changed with time, as in experiments recently proposed to detect the dynamical Casimir effect \cite{Agnesi09,Antezza14}.

We now consider the case of an open JC system, in which either the transition frequency of the two-level system or the frequency of the cavity field mode periodically change with time.
As mentioned above, the first case could be obtained by subjecting the atom to an external laser field which periodically modulates its energy levels by time-dependent Stark shift, while
the second case could be obtained by periodically moving one of the walls of a one-dimensional cavity. Both cases allow energy gain or loss for the coupled system.
In the first case, the JC Hamiltonian becomes
\begin{equation}
\label{JCHt2}
H = \hbar \wo (t) \left( \bd b -\frac 12 \right) + \hbar \omega \ad a + g \left( a\bd + \ad b \right) \, ,
\end{equation}
while in the second one we have
\begin{equation}
\label{JCHt1}
H = \hbar \wo \left( \bd b -\frac 12 \right) + \hbar \omega (t) \ad a + g \left( a\bd + \ad b \right) \, ,
\end{equation}
where $\omega (t)$ and $\wo (t)$ are prescribed functions of time that will be specified later on. We assume that the coupling constant $g$ is real, and thus the time-dependent Hamiltonians \eqref{JCHt1} and \eqref{JCHt2} are Hermitian operators.
A non-adiabatic modulation of interacting atom-field parameters has been also recently recognized as a way to obtain new phenomena {in quantum electrodynamics (QED)} involving photon-polaritons coupling \cite{Gunter09}, dynamical Casimir-Polder forces \cite{VP08,MVP10} and generation of quantum vacuum radiation \cite{LGCC09,Law13}.
Cases somehow analogous to the present one have been recently considered for the Lee model \cite{LV12}, thus restricted to the one-excitation subspace only. In our case, they are generalized to an arbitrary $n$-excitation subspace, including the strong coupling regime near resonance, although for a single cavity mode.

\section{Solution of the Modulated Jaynes-Cummings Model}
\label{sec:JCSolution}

We now write the equations of motion for the state amplitudes in a generic $n$-excitation subspace for our modulated Jaynes-Cummings model, for both cases of periodically driven atomic transition frequency or field-mode frequency, described by Hamiltonians \eqref{JCHt1} and \eqref{JCHt2} respectively. In both cases, in the $n$-excitations subspace, we can write the state in the following form
\begin{equation}
\label{state-n-excitations}
\mid \psi_n (t) \rangle = \tilde{c}'(t) \mid n-1, 1 \rangle +  d'(t) \mid n, 0 \rangle \, ,
\end{equation}
where, {as we said before}, the first element in the states refers to the bosonic field and the second one to the fermionic field. Substituting \eqref{state-n-excitations} into the Schr\"{o}dinger equation, we obtain a set of differential equations for the coefficients $\tilde{c}'(t)$ and $d'(t)$.

In the case of a time-varying atomic transition frequency, using the Hamiltonian {\eqref{JCHt2}}, we obtain
\begin{eqnarray}
\label{eqdiff-coeff-c}
i \frac d{dt}\tilde{c}'(t) &=& \left[ \frac{\wo (t)}2 +\omega (n-1) \right] \tilde{c}'(t) + \frac {g\sqrt{n}} \hbar  d'(t) \, , \\
\label{eqdiff-coeff-d}
i \frac d{dt}d'(t) &=& \left[ n\omega - \frac{\wo (t)}2 \right] d'(t) + \frac {g\sqrt{n}}\hbar  \tilde{c}'(t) \, .
\end{eqnarray}

We now assume that the frequency $\wo (t)$ oscillates periodically in time according to
\begin{equation}
\label{Omega0-t}
\wo (t) = \wo \left[ 1+ \beta \cos (\Omega t) \right] \, ,
\end{equation}
where $\omega_0$ is the average frequency, $\beta$ is a complex parameter and $\Omega$ is the frequency of the external modulation. From a physical point of view, a complex $\beta$ can be thought as a gain or loss term for the atomic energy, as we will discuss in more detail later on.

We now define
\begin{eqnarray}
\tilde{c}'(t) &\equiv& e^{-i\left[ \frac {\wo}2 + \omega (n-1) \right] t} \, \tilde{c}(t) \, , \label{int-repr-1} \\
d'(t) &\equiv& e^{-i\left( n\omega - \frac {\wo}2 \right) t} \, d(t) \, . \label{int-repr-2}
\end{eqnarray}
This is equivalent to using the interaction representation with an atomic Hamiltonian where the average atomic frequency $\omega_0$ appears. Equations \eqref{eqdiff-coeff-c} and \eqref{eqdiff-coeff-d} thus yield
\begin{eqnarray}
\label{eqdiff-coeff-c1}
i \frac d{dt}\tilde{c}(t) &=& \frac {\wo \beta}2 \cos (\Omega t) \tilde{c}(t) + \frac {g\sqrt{n}}\hbar e^{i(\wo -\omega )t} d(t) \, , \\
\label{eqdiff-coeff-d1}
i \frac d{dt}d(t) &=& -\frac {\wo \beta}2 \cos (\Omega t) d(t) + \frac {g\sqrt{n}}\hbar e^{-i(\wo -\omega )t}  \tilde{c}(t) \, .
\end{eqnarray}

We also define
\begin{equation}
\label{ct}
c (t) \equiv \tilde{c}(t) e^{-i \frac {\wo \beta}{2\Omega} \sin (\Omega t)}  \, .
\end{equation}

Using definitions \eqref{int-repr-1}, \eqref{int-repr-2} and \eqref{ct}, equations \eqref{eqdiff-coeff-c1} and  \eqref{eqdiff-coeff-d1} for the amplitudes give
\begin{eqnarray}
i \frac d{dt}c(t) &=& \frac {g \sqrt{n}}\hbar e^{i(\wo -\omega )t}e^{i \frac{\wo \beta}{2\Omega} \sin (\Omega t)} d(t), \label{eq-c} \\
i \frac d{dt}d(t) &=& -\frac {\wo \beta}2 \cos (\Omega t) d(t) \nonumber \\
&+& \frac {g \sqrt{n}}\hbar e^{-i(\wo -\omega )t} e^{-i\frac{\wo \beta}{2\Omega} \sin (\Omega t)} c(t) \label{eq-d} \, .
\end{eqnarray}

In the absence of external modulation of the atomic frequency, that is $\beta =0$, and near resonance $\wo \simeq \omega$, the solution of the equations above is an oscillation of the amplitudes at the Rabi frequency $\Omega_R =g\sqrt{n}/\hbar$.
When the modulation given by \eqref{Omega0-t} is acting on the system with $\Omega \gg \Omega_R$, a reasonable assumption because the Rabi frequency is usually much smaller than typical atomic frequencies \cite{KGKKS06}, the amplitudes $c(t)$ and $d(t)$ are expected to evolve much more slowly than changes given by the modulation frequency $\Omega$.
In this case we can average Equations \eqref{eq-c} and \eqref{eq-d} over a period relative to the modulation frequency $\Omega$, similarly as it is done in \cite{LV12}, bringing slowly-varying terms outside the time integrals. Using
\begin{equation}
\label{average}
\frac 1{2\pi}\int_{-\pi}^\pi e^{-i\Gamma \sin x}dx = J_0(\Gamma ) \, ,
\end{equation}
where $\Gamma$ is a complex number and $J_0(z)$ is a Bessel function \cite{AS65}, and observing that $J_0(z)$ is an even function, the averaged equations at resonance $\wo \simeq \omega$ are thus
\begin{eqnarray}
i \frac d{dt}c(t) &=& \frac {g\sqrt{n}} \hbar J_0\left(-\frac{\wo \beta}{2\Omega}\right) d(t) \, , \label{eq-c1} \\
i \frac d{dt}d(t) &=& \frac {g\sqrt{n}} \hbar  J_0\left(-\frac{\wo \beta}{2\Omega}\right) c(t)
\label{eq-d1} \, .
\end{eqnarray}

As a specific case, if we put $n=1$ in our results we recover previous results in \cite{LV12} for the one-excitation subspace (Lee-Friedrichs model).

We can now compare Eqs. \eqref{eq-c1} and \eqref{eq-d1} with the analogous equations obtained in the static case ($\beta =0$), that is a Jaynes-Cummings model at resonance with time-independent atomic and field frequencies, described by the Hamiltonian \eqref{JCHamiltonian}
\begin{eqnarray}
i \frac d{dt}c(t) &=& \frac {g_s\sqrt{n}} \hbar d(t) \, , \label{eq-c1s} \\
i \frac d{dt}d(t) &=& \frac {g_s\sqrt{n}} \hbar c(t)
\label{eq-d1s} \, .
\end{eqnarray}

If we choose the complex parameter $\beta$ in such a way that
\begin{equation}
J_0\left(-\frac{\wo \beta}{2\Omega}\right)=i \, ,
\label{beta}
\end{equation}
equations \eqref{eq-c1} and \eqref{eq-d1} for the modulated case coincide with those for the static case given by Eqs. \eqref{eq-c1s} and \eqref{eq-d1s} with
\begin{equation}
g_s = ig  \in {\Bbb C} \, ,
\label{imcoupling}
\end{equation}
that is with an imaginary coupling constant making non-Hermitian the (time-independent) Hamiltonian \eqref{JCHamiltonian}. We point out that equation \eqref{beta} has a complex solution given by $\frac{\wo \beta}{2\Omega} \simeq -2.14+1.42 \, i$, yielding a complex value of $\beta$.

Same considerations apply to the case of a modulated field frequency $\omega (t)$, described by the Hamiltonian {\eqref{JCHt1}}. Also in this case we may assume a periodic modulation of the frequency of the cavity mode,
\begin{equation}
\omega (t) = \omega \left( 1+ \beta \cos \Omega t \right) \, ,
\label{Omega-t}
\end{equation}
with $\beta$ a complex parameter. Using \eqref{state-n-excitations}, we can write the equations for the amplitudes and use definitions \eqref{int-repr-1} and \eqref{int-repr-2}. Defining now
\begin{equation}
\label{ct-2}
c (t) \equiv \tilde{c}(t) e^{-i \frac {n\omega \beta}\Omega \sin (\Omega t)}  \, ,
\end{equation}
we obtain a set of differential equations  for the amplitudes analogous to \eqref{eq-c1} and \eqref{eq-d1}.
Thus, also in this case, comparison with Eqs. \eqref{eq-c1s} and \eqref{eq-d1s} shows that by the an appropriate choice of the system parameters such that
\begin{equation}
J_0\left(-\frac{n\omega \beta}\Omega\right)=i \, ,
\label{beta1}
\end{equation}
we can simulate, also in this case, the Hermitian JC Hamiltonian with a modulated cavity frequency by a static non-Hermitian JC Hamiltonian with an imaginary coupling constant.

A complex value of the parameter $\beta$ in \eqref{Omega0-t} as obtained by the solutions of equations \eqref{beta} and \eqref{beta1}, could result from several microscopic physical mechanisms involving energy exchange from or to the system, for example from the interaction of the atom with an external environment whose properties are modulated in time. Such a modulated environment could be a set of infinite harmonic oscillators with a continuous energy spectrum in out of equilibrium conditions (that can exchange energy with the atom), by a modulated (i.e. oscillating) optical cavity or by a dynamical photonics crystal.

These results thus give a physical meaning to the non-Hermitian Jaynes-Cummings Hamiltonian \eqref{JCHamiltonian} with an imaginary coupling constant $g_s = ig$, $g$ being a real quantity.
In other words, our results clearly show that a Jaynes-Cummings model with an imaginary coupling constant, yielding a non-Hermitian Hamiltonian, can be a useful model to simulate a realistic Hermitian Jaynes-Cummings system with some external modulation of the physical parameters, specifically the cavity mode frequency or the atomic transition frequency, at least when gain-loss effects want to be efficiently described. This gives a physical support to further investigate a non-Hermitian Jaynes-Cummings model, from both physical and mathematical aspects.

In recent years, it has been discussed that physical systems described by non-Hermitian ${\mathcal PT}$ symmetric Hamiltonians can be investigated in the framework of the more general formalism  of pseudo-Hermitian operators \cite{Mostafazadeh10, BAMY04}. As we shall discuss in the next Section, pseudo-operators are defined by introducing a particular deformed version of the canonical commutation relations. The meaning of the pseudo-operators and their relation with ${\mathcal PT}$ symmetry has been recently considered, both from a purely mathematical point of view and also in view of possible applications to the so-called pseudo-Hermitian quantum mechanics, \cite{bagbookall}. Pseudo-Hermitian quantum mechanics has received a lot of interest since the discovery of the fact that the well known cubic Hamiltonian $H=p^2+ix^3$, has purely real and discrete spectrum \cite{Bender07}.
The interest on this subject is also motivated by the fact that the spectral properties of ${\mathcal PT}$ symmetric Hamiltonians follow from their pseudo-Hermiticity \cite{Mostafazadeh10}.

The relevance of pseudo-Hermitian quantum mechanics in the framework of quantum optics and open quantum systems has been discussed. For example, it has been recently shown that non-Hermitian interaction between atoms and the electromagnetic field can be appropriately investigated by exploiting the methods of pseudo-Hermitian quantum mechanics \cite{BAMY04}.

In the next section,  on the basis of the previous results showing the physical interest of the non-Hermitian Jaynes-Cummings Hamiltonian with a purely imaginary coupling constant, we will discuss a deformed version of the Jaynes-Cummings Hamiltonian, in which the bosonic and fermionic modes are replaced by their pseudo-bosonic and pseudo-fermionic versions, \cite{bagbook}. This interest is motivated by several recent results, see for instance \cite{BL13,FBbagpb4,FBgarg} and  \cite{bagbook} for a recent review, where many systems introduced along the years in connection with ${\mathcal PT}$ or pseudo-symmetric Hamiltonians, where complex-valued parameters are introduced, have been shown to be expressible in terms of pseudo-bosonic or pseudo-fermionic operators. In doing so, the extra bonus one gets is that the eigensystems of these Hamiltonians, and of their adjoint, can be easily deduced, following a somehow standard procedure, procedure which will be used to diagonalize the Hamiltonian in \eqref{21} below,  generalizing to the case of pseudo-operators the procedure used in the Hermitian case. This will have also very interesting mathematical consequences, as we will see, since pseudo-bosonic and pseudo-fermionic structures are {\em mixed} along the way.
Then we conclude that the deformed model of the Jaynes-Cummings Hamiltonian we introduce below is thus physically well-motivated and permits one to appropriately investigate the non-Hermitian interaction between a two-level system and the electromagnetic field in a cavity.

\section{The non-Hermitian Jaynes-Cummings Hamiltonian}
\label{sec:nonHJC}
{Pseudo-bosons and pseudo-fermions are defined performing a particular  {\em deformation} of the canonical commutation relations for bosons and fermions. In particular,
pseudo-bosons arise from the canonical commutation relation $[a,a^{\dagger}]=1$ after replacing $a^{\dagger}$ by another operator $A$ satisfying the relation $[a,A]=1$, with $A$ in general different from $a^{\dagger}$.  More recently, a similar procedure to define pseudo-fermions has been carried out, and their properties have been analyzed again both from a mathematical and from a physical point of view \cite{bagbook}.
We now introduce our model for a non-Hermitian Jaynes-Cummings Hamiltonian expressed in terms of pseudo-bosonic and pseudo-fermionic operators, and we discuss its diagonalization.}
\subsection{The deformed Jaynes-Cummings Hamiltonian}
\label{ssec:defJC}

The Hamiltonian of our model, which can be considered as the extension of the Hamiltonian \eqref{JCHamiltonian} considered in Section \ref{sec:JCModel},  can be written as
\begin{equation}
H_\alpha=\hbar\omega_0\left(C_\alpha c_\alpha-\frac{1}{2}\right)+\hbar\omega D_\alpha d_\alpha+\epsilon d_\alpha C_\alpha+\epsilon^* D_\alpha c_\alpha \, ,
\label{21}
\end{equation}
acting on a Hilbert space $\Hil :=\Hil_b\otimes\Hil_f$, where $\Hil_f={\Bbb C}^2$ (fermionic sector) while $\Hil_b$ is infinite dimensional (bosonic sector), $\epsilon$ indicates the coupling constant (we are using different symbols in order to avoid confusion between the {Hermitian} and non-Hermitian cases). The suffix $\alpha$ is a parameter which, when is sent to zero (in an appropriate way) produce a self-adjoint Hamiltonian $H_0=H_0^\dagger$. On the other hand, since for $\alpha\neq 0$ we have, in general $c_\alpha^\dagger\neq C_\alpha$ and $d_\alpha^\dagger\neq D_\alpha$, we expect that $H_\alpha\neq H_\alpha^\dagger$. More in details, following a general result discussed in \cite{bagbook}, what we have in mind is that $c_\alpha$, $C_\alpha$, $d_\alpha$ and $D_\alpha$ are (at least formally) similar to two pairs of bosonic ($c$ and $c^\dagger$) and fermionic ($d$ and $d^\dagger$) operators, and that this similarity is implemented by some invertible operator $Q_\alpha$, possibly unbounded with unbounded inverse, such that $Q_\alpha$ converges to the identity operator (in some suitable topology) when $\alpha$ goes to zero. We will briefly return to this point in Section \ref{ssec:sim}.

We assume here the following rules
\begin{eqnarray}
[d_\alpha\otimes\1_f,D_\alpha\otimes\1_f]=\1_b\otimes\1_f=:\1 \, , \nonumber \\
\qquad \{\1_b\otimes c_\alpha,\1_b\otimes C_\alpha\}=\1 \, ,
\label{22}
\end{eqnarray}
while all the other commutators are zero. Hence the pair $(d_\alpha,D_\alpha)$ satisfies the pseudo-bosonic commutation rules, while $(c_\alpha,C_\alpha)$ behaves as pseudo-fermions.

With a simple extension of the procedure discussed in Section \ref{sec:JCModel} (see also \cite{CPP95}), we can rewrite $H_\alpha$ in a diagonal form. For that we first introduce a global non self-adjoint number operator, analogous to the excitation-number operator \eqref{NumberExc},
\begin{equation}
N_\alpha=D_\alpha d_\alpha+C_\alpha c_\alpha \, ,
\label{23}
\end{equation}
and a map $T_\alpha$ defined as follows:
\begin{equation}
T_\alpha=\exp\left\{-\theta_\alpha(4|\epsilon|^2N_\alpha)^{-1/2}\left(\epsilon d_\alpha C_\alpha-\epsilon^* D_\alpha c_\alpha\right)\right\} \, ,
\label{24}
\end{equation}
where $\theta_\alpha$ is an operator defined, in analogy with (\ref{ThetaOperator}), via the conditions $\sin\theta_\alpha=-(4|\epsilon|^2N_\alpha)^{1/2}\Delta_\alpha^{-1}$ and $\cos\theta_\alpha=-\delta\Delta_\alpha^{-1}$, where $\delta=\hbar(\omega_0-\omega)$ is the detuning between the energies of the two fields, and $\Delta_\alpha=\left(\delta^2+4|\epsilon|^2N_\alpha\right)^{1/2}$, which is clearly invertible, apart from when restricted to the zero-excitation subspace (we are not however interested in this subspace, because our Hamiltonian is already diagonal in this subspace).

It is easy to see that the new operators $\hat d_\alpha\otimes\hat \1_f=T_\alpha(d_\alpha\otimes\1_f)T_\alpha^{-1}$, $\hat D_\alpha\otimes\hat \1_f=T_\alpha(D_\alpha\otimes\1_f)T_\alpha^{-1}$, $\hat 1_b\otimes\hat c_\alpha=T_\alpha(\1_b\otimes c_\alpha)T_\alpha^{-1}$ and $\hat 1_b\otimes\hat C_\alpha=T_\alpha(\1_b\otimes C_\alpha)T_\alpha^{-1}$ still obey the same rules as the operators without the hat: hence they are tensor products of pseudo-bosonic and pseudo-fermionic operators. Most important, in terms of these operators the Hamiltonian $H_\alpha$ turns out to be in a diagonal form
\begin{equation}
H_\alpha=\left(\hbar\omega-\hat\Delta_\alpha\right)\left(\hat C_\alpha\hat c_\alpha-\frac{1}{2}\right)+ \hbar\omega\hat D_\alpha\hat d_\alpha \, ,
\label{25}
\end{equation}
where $\hat\Delta_\alpha=\left(\delta^2+4|\epsilon|^2\hat N_\alpha\right)^{1/2}=\Delta_\alpha$, since $\hat N_\alpha=N_\alpha$.

The eigenvectors of $H_\alpha$ (and of $H_\alpha^\dagger$) can now easily computed adapting to the present situation the general framework of deformed canonical commutation relations and canonical anti-commutation relations discussed in \cite{bagbook}. We first assume that two non-zero vectors $\hat\varphi_0$ and $\hat\psi_0$ do exist in $\Hil_b$ such that, if $\hat\eta_0$ and $\hat\mu_0$ are two vectors of fermionic Hilbert space $\Hil_f$ annihilated respectively by $\hat c_\alpha$ and $\hat C_\alpha^\dagger$, we have
\begin{equation}
\left(\hat d_\alpha\otimes\hat \1_f\right)\hat\Phi_{0,0}=\left(\hat 1_b\otimes\hat c_\alpha\right)\hat\Phi_{0,0}=0,
\label{26}
\end{equation}
as well as
\begin{equation}
\left(\hat D_\alpha^\dagger\otimes\hat \1_f\right)\hat\Psi_{0,0}=\left(\hat 1_b\otimes\hat C_\alpha^\dagger\right)\hat\Psi_{0,0}=0 \, ,
\label{27}
\end{equation}
where $\hat\Phi_{0,0}:=\hat\varphi_0\otimes\hat\eta_0$ and $\hat\Psi_{0,0}:=\hat\psi_0\otimes\hat\mu_0$.

Following \cite{bagbook}, it is convenient to assume that $\hat\varphi_0$ and $\hat\psi_0$ belong to a dense domain $\D$ of $\Hil_b$, which is stable under the action of $d_\alpha$, $D_\alpha$, and their adjoint. Of course, nothing has to be required to the fermionic operators and to the related vectors, since $\Hil_f$ is a finite dimensional vector space.

If such a $\D$ exists, then we can use the two vacua $\hat\Phi_{0,0}$ and $\hat\Psi_{0,0}$ to construct two different set of vectors, $\F_{\hat\Phi}:=\{\hat\Phi_{n,k}, \,n\geq0, \, k=0,1\}$ and
$\F_{\hat\Psi}:=\{\hat\Psi_{n,k}, \,n\geq0, \, k=0,1\}$, all belonging to $\D\otimes\Hil_f$, as follows:
\begin{eqnarray}
\hat\Phi_{n,k}&=&\left(\frac{1}{\sqrt{n!}}\hat D_\alpha^n\otimes \hat C_\alpha^k\right)\hat\Phi_{0,0}
\nonumber \\
&=&\left(\frac{1}{\sqrt{n!}}\hat D_\alpha^n\hat\varphi_0\right)\otimes\left(\hat C_\alpha^k\hat\eta_0\right)=:\hat\varphi_n\otimes\hat\eta_k \, \
\label{28}
\end{eqnarray}
and
\begin{eqnarray}
\hat\Psi_{n,k}&=&\left(\frac{1}{\sqrt{n!}}\hat {d_\alpha^\dagger}^n\otimes \hat {c_\alpha^\dagger}^k\right)\hat\Psi_{0,0}
\nonumber \\
&=&\left(\frac{1}{\sqrt{n!}}\hat {d_\alpha^\dagger}^n\hat\psi_0\right)\otimes\left(\hat {c^\dagger}_\alpha^k\hat\mu_0\right)=:\hat\psi_n\otimes\hat\mu_k \, ,
\label{29}
\end{eqnarray}
with obvious notation, where $n=0,1,2,\ldots$ and $k=0,1$. It is now easy to check that
\begin{equation}
H_\alpha\hat\Phi_{n,k}=E_{n,k}\hat\Phi_{n,k},\qquad H_\alpha^\dagger\hat\Psi_{n,k}=E_{n,k}\hat\Psi_{n,k} \, ,
\label{210}
\end{equation}
where
\begin{equation}
E_{n,k}=\left[ \hbar\omega-\left(\delta^2+4|\epsilon|^2(n+k+1)\right)^{1/2}\right]\left(k-\frac{1}{2}\right)+\hbar\omega n \, .
\label{211}
\end{equation}

Incidentally, we see that $E_{n,0}$ is surely bounded from below in $n$, independently of the choice of the parameters. As for $k=1$, we see that this is true only if $\hbar\omega\geq2|\epsilon |$, which is therefore to be assumed to give our Hamiltonian a physical meaning.

Equation \eqref{210} ensures us that $\F_{\hat\Phi}$ and
$\F_{\hat\Psi}$ are biorthogonal. Indeed, if the normalizations of $\hat\Phi_{0,0}$ and $\hat\Psi_{0,0}$ are chosen in such a way that $\left<\hat\Phi_{0,0},\hat\Psi_{0,0}\right>=1$, then
\begin{equation}
\left<\hat\Phi_{n,k},\hat\Psi_{m,l}\right>=\left<\hat\varphi_{n},\hat\psi_{m}\right>_{\Hil_b}\left<\hat\eta_{k},\hat\mu_{l}\right>_{\Hil_f}=\delta_{n,m}\delta_{l,k} \, .
\label{212}
\end{equation}
Here $\left<.,.\right>_{\Hil_b}$ and $\left<.,.\right>_{\Hil_f}$ are respectively the scalar products in $\Hil_b$ and in $\Hil_f$.

A serious problem, as it happens quite often in this kind of problems, is whether $\F_{\hat\Phi}$ and
$\F_{\hat\Psi}$ are bases of $\Hil$ or not. Or, at least, if they are $\G$-quasi bases, see below, for some $\G$ dense in $\Hil$. In general, the answer is hard to find. However, there is at least one situation where this can be done, as we are going to discuss next.

\subsection{What if $H_\alpha$ is similar to a self-adjoint operator}
\label{ssec:sim}

We consider now the following self-adjoint Hamiltonian
\begin{equation}
H_0=\hbar\omega_0\left(c^\dagger c-\frac{1}{2}\right)+\hbar\omega d^\dagger d+\epsilon d c^\dagger+\epsilon^* d^\dagger c \, ,
\label{213}
\end{equation}
where $[d,d^\dagger]=\1_b$, $\{c,c^\dagger\}=\1_f$, $c^2=0$, and all the other commutators are zero. It is clear that $H_0=H_0^\dagger$. It is also evident that this can be obtained from $H_\alpha$ in \eqref{21} replacing pseudo-bosonic (fermionic) operators with their {\em standard} counterparts or, in view of our previous comment, sending $\alpha$ to zero. $H_0$ is the Jaynes-Cummings Hamiltonian essentially discussed in \cite{JC63,CPP95}, and it can be diagonalized as we did for $H_\alpha$. In particular, introducing new bosonic and fermionic operators by means of the unitary operator $T_0$, see \eqref{24}, the eigenstates of $H_0$ can be easily deduced. More explicitly, introducing $\hat d\otimes\hat \1_f=T_0(d\otimes\1_f) {T_0^{-1}}$, and $\hat 1_b\otimes\hat c=T_0(\1_b\otimes c)T_0^{-1}$, the Hamiltonian can be rewritten as
\begin{equation}
H_0=\left(\hbar\omega-\hat\Delta_0\right)\left(\hat c^\dagger\hat c-\frac{1}{2}\right)+ \hbar\omega\hat d^\dagger\hat d \, ,
\label{214}
\end{equation}
where $\hat\Delta_0=\left(\delta^2+4|\epsilon|^2\hat N_0\right)^{1/2}=\Delta_0$, since $\hat N_\alpha=N_\alpha$.

The eigenstates $\hat\Phi_{n,k}^o$ of $H_0$ can be defined easily: first we introduce the vacua of $\hat d$ and $\hat c$, $\hat \varphi_0^o\in\Hil_b$ and $\hat \eta_0^o\in\Hil_f$: $\hat d\hat \varphi_0^o=\hat c\hat \eta_0^o=0$. These vectors surely exist, as it is clear. Then we construct, more solito, $\hat \varphi_n^o=\frac{1}{\sqrt{n!}}\,(\hat d^\dagger)^n\hat \varphi_0^o$ and $\hat \eta_1^o=\hat c^\dagger\hat \eta_0^o$, and finally we define $\hat \Phi_{n,k}^o=\hat \varphi_n^o\otimes\hat \eta_k^o$, $n\geq0$ and $k=0,1$. Since $H_0\hat \Phi_{n,k}^o=E_{n,k}\hat \Phi_{n,k}^o$, $H_0$, $H_\alpha$ and $H_\alpha^\dagger$ are all isospectral. Hence intertwining operators between them are expected to exist, \cite{intop}. For this reason, it is not a strong assumption to assume here that a self-adjoint operator $S_\alpha=S_\alpha^\dagger$ does exist, at least on some dense subset of $\Hil$, $ H_\alpha S_\alpha=S_\alpha H_0$. Then, if we further assume that $S_\alpha$ is invertible, we see that $H_\alpha=S_\alpha H_0 S_\alpha^{-1}$. Of course we also have $H_\alpha^\dagger=S_\alpha^{-1} H_0 S_\alpha$ and, more important for us,
\begin{equation}
\hat\Phi_{n,k}=S_\alpha \Phi_{n,k}^o,\qquad \hat\Psi_{n,k}=S_\alpha^{-1} \Phi_{n,k}^o \, ,
\label{215}
\end{equation}
at least if $\Phi_{n,k}^o\in D(S_\alpha)\cap D(S_\alpha^{-1})=:\G$, where $D(S_\alpha)$ is a dense domain where the operator $S_\alpha$ is defined, and if the multiplicity of each $E_{n,k}$ is one.
These equations are in agreement with formula \eqref{212}; this is because the set $\F_{\hat\Phi^o}=\{\Phi_{n,k}^o\}$ is an orthonormal basis for $\Hil$. But we get more than this: in fact, if $S_\alpha$ and $S_\alpha^{-1}$ are both bounded, then $\F_{\hat\Phi}$ and $\F_{\hat\Psi}$ are biorthogonal Riesz bases, which is the best we can have (from a technical point of view) when we lose orthonormality, \cite{you,heil,chri}. However, when $S_\alpha$ or $S_\alpha^{-1}$, or both, are not bounded, the situation is not so simple and, in fact, only some weak resolutions of the identity can be deduced \cite{bagbook}. In particular,  assuming that $\G$ is dense in $\Hil$, we can conclude that $\F_{\hat\Phi}$ and $\F_{\hat\Psi}$ are biorthogonal $\G$-quasi bases. We refer to \cite{bagbook} for more information about the properties of these sets. Indeed, they give rise to several interesting properties which have been investigated in recent years. Here we just want to mention that, in the present context, when we say that $\F_{\hat\Phi}$ and $\F_{\hat\Psi}$ are $\G$-quasi bases we mean that, taken two arbitrary vectors $f$ and $g$ in $\G$, the following equalities are satisfied:
\begin{equation}
\left<f,g\right>=\sum_{n,k}\left<f,\hat\Phi_{n,k}\right>\left<\hat\Psi_{n,k},g\right>=\sum_{n,k}\left<f,\hat\Psi_{n,k}\right>\left<\hat\Phi_{n,k},g\right> \, .
\end{equation}

From \eqref{215} we also deduce that
\begin{equation}
\hat\Phi_{n,k}=S_\alpha^2\hat\Psi_{n,k},
\label{216}
\end{equation}
so that $\hat\Psi_{n,k}=S_\alpha^{-2}\hat\Phi_{n,k}$. These equations show that (i) a {\em metric} operator, see \cite{bagbook} can be introduced in the game, at least formally, which is positive, self-adjoint, invertible, and maps one set of eigenstates into the other; (ii) the following equalities hold, on some domain
\begin{equation}
S_\alpha^2=\sum_{n,k}|\hat\Phi_{n,k}\left>\right<\hat\Phi_{n,k}| \, , \,\,\,\, S_\alpha^{-2}=\sum_{n,k}|\hat\Psi_{n,k}\left>\right<\hat\Psi_{n,k}| \, .
\end{equation}
Here $\left(|f><g|\right)h:=<g,h>f$, for all $f,g,h\in\Hil$. More mathematical details on the general framework are discussed in \cite{bagbook}, to which we refer the interested reader.

From a physical point of view, the advantage of the diagonalization of the non-Hermitian Jaynes-Cummings Hamiltonian obtained in this section through the operator \eqref{24}, in terms of pseudo-bosonic and pseudo-fermionic operators, is that it allows us to have an explicit form of the diagonal Hamiltonian and of the dressed states, that can be obtained from the bare ones by applying the operator  \eqref{24}, in analogy with the usual Hermitian case. This can be helpful in investigating many physical systems that can be simulated by a non-Hermitian Hamiltonian, such as the system considered in Section \ref{sec:JCSolution} , modulated optical lattices and the Lee model in the ghost regime \cite{LV12,LV14}.

\section{Conclusions}
\label{sec:Concl}

In this paper we have considered two time-dependent Jaynes-{Cummings}-type Hamiltonian models, describing a two-level system interacting with a single cavity mode. More in details, we have considered  a periodically driven transition frequency of the two-level system and a periodic change of the frequency of the cavity mode. These two interesting situations could be obtained, respectively, by a dynamic Stark shift of the energy levels of an atom by a laser field, and by a motion of the conducting wall of a cavity or even a dynamical mirror. We have first shown that, for appropriate choices of the system parameters, both cases can be simulated with a static non-Hermitian ${\mathcal PT}$ symmetric JC Hamiltonian with an imaginary coupling constant. This result suggests the relevance, even from a physical point of view,  to consider  the non-Hermitian JC model. {With this in mind, we have analyzed our system by exploiting a more general procedure for treating non-Hermitian Hamiltonians. We have introduced a model of {\em deformed} Jaynes-Cummings Hamiltonian expressed in term of pseudo-operators,} and extended the well-known unitary operator diagonalization of the Hermitian JC Hamiltonian with a invertible operator to the non-Hermitian case in terms of pseudo-bosonic and pseudo-fermionic operators. This has allowed us to obtain explicitly its diagonal form and its eigenstates and eigenvalues, analogously to the known Hermitian case. {This has also very important consequences from the mathematical point of view, since pseudo-bosonic and pseudo-fermionic structures are {\em  shuffled} along the way.} Mathematical and physical implications and applications of our results have been also discussed in detail.
{The {\em deformed} JC model introduced could be used to further investigate the interaction between atomic systems and the electromagnetic field, including damping or amplifying processes, which is of fundamental importance for example in quantum optics.}

\section*{ACKNOWLEDGMENTS}
Financial support by the Julian Schwinger Foundation, MIUR, University of Palermo, GNFM and CRRNSM is gratefully acknowledged.

\end{document}